# Persistent Topological Surface State at the Interface of $Bi_2Se_3$ Film Grown on Patterned Graphene

Namdong Kim,[*,†,‡,∥] Paengro Lee,[‡] Youngwook Kim,[‡] Jun Sung Kim,[‡] Yongsam Kim,[§,∥] Do Young Noh,[§] Seong Uk Yu,[†] Jinwook Chung,[‡] and Kwang S. Kim[*,†,‡,#]

[†]Department of Chemistry, [‡]Department of Physics, [∥]Pohang Accelerator Laboratory, Pohang University of Science and Technology, Pohang 790-784, Korea, [§]Department of Physics and Photon Science, Gwangju Institute of Science and Technology, Gwangju 500-712, Korea, and [#]Department of Chemistry, Ulsan National Institute of Science and Technology, Ulsan 689-798, Korea

*To whom correspondence should be addressed to K.S.K. (e-mail: kimks@unist.ac.kr) or N.K. (email: east@postech.ac.kr).

**ABSTRACT**  We employed graphene as a patternable template to protect the intrinsic surface states of thin films of topological insulators (TIs) from environment. Here we find that the graphene provides high-quality interface so that the Shubnikov de Haas (SdH) oscillation associated with a topological surface state could be observed at the interface of a metallic $Bi_2Se_3$ film with a carrier density higher than $\sim 10^{19}$ cm$^{-3}$. Our *in situ* X-ray diffraction study shows that the $Bi_2Se_3$ film grows epitaxially in a quintuple layer-by-layer fashion from the bottom layer without any structural distortion by interfacial strain. The magnetotransport measurements including SdH oscillations stemming from multiple conductance channels reveal that the topological surface state, with the mobility as high as $\sim 0.5$ m$^2$/Vs, remains intact from the graphene underneath without degradation. Given that the graphene was prepatterned on arbitrary insulating substrates, the TI-based microelectronic design could be exploited. Our study thus provides a step forward to observe the topological surface states at the interface without degradation by tuning the interface between TI and graphene into a measurable current for device application.

**KEYWORDS**   Topological insulator (TI), Surface states, $Bi_2Se_3$ thin film, Graphene, Interface

Topological insulators (TIs) are a new class of three-dimensional insulating materials with metallic surface channels caused by strong spin-orbit interaction.[1,2] With the discovery of quantum spin Hall effect in two-dimensional HgTe/(Hg,Cd)Te quantum wells,[3] three-dimensional TI systems have been observed by angle-resolved photoemission spectroscopy (ARPES) after the first realization in $Bi_{1-x}Sb_x$.[4] Despite their unique and exotic properties,[5–7] unlike graphene,[8] the application of TI materials has been huddled due essentially to the dominating bulk conduction and rapid degradation of surface channels during fabrication process.[9–11] Preserving the unique surface states of TIs has been one of the unresolved issues to utilize TI materials in device applications to ensure the presence of surface states not hampered by the dominant bulk conduction, and well protected from any contamination in ambient atmosphere.[9,12] Such a disadvantage has limited device applications and, in particular, an overwhelming bulk conduction makes transport measurements of the surface states difficult.[9,13] For example, Shubnikov de Haas (SdH) oscillations of the surface states were barely observed at huge magnetic fields in previous attempts to reduce bulk carriers by Sb doping.[9] However, such observations were successful only for bulk-insulating samples with low carrier densities ($< \sim 10^{17}$ $cm^{-3}$).[9,13] In another way, minimizing interfacial interactions is of utmost importance in realizing enhanced performance of tuning the intrinsic surface states of TI thin films,[14] as already reported in the heterostructural system such as graphene/$h$BN[15] and $Bi_2Te_2Se$/$h$BN[16] exhibiting higher mobilities. Here, we present a pragmatic way to resolve such huddles by forming a buried interface of as-grown $Bi_2Se_3$ upon a graphene-based predesigned device layer (Figure 1). We find that our surface states stemming from the buried interface between TI and graphene remain intact, easily detectable despite the bulk conduction, and are apparently well protected also from environmental contamination. We therefore provide an efficient and practical means to utilize surface states of TI in device applications. Thin TI film grown on a patterned graphene can be fully utilized to meet the demands for a large-scale integration in device applications.

**RESULTS AND DISCUSSION**

Since it is essential to minimize the interface interactions to preserve the surface state of TI by forming a well-defined interface, we carried out X-ray diffraction measurements to examine the morphology of the interface formed between graphene and TI. We have first prepared different types of large-area graphene samples for an *in situ* X-ray diffraction: (i) chemical vapour deposition (CVD) graphene on $SiO_2$ and (ii) multilayer graphene (MLG) grown on 6H–SiC(0001) surface.[17] Film growth has been performed by thermally evaporating small pieces of crystal with substrate maintained near 300 °C. Layered $Bi_2Se_3$ films are epitaxially grown with sufficient thermal energies to allow dynamic atomic motion on surface under effective growth rate.[18–20] We confirm that a single-layer graphene (SLG) remains intact underneath the $Bi_2Se_3$ film by observing G, 2D bands (1588, 2680 cm$^{-1}$) corresponding to SLG in addition to the characteristic peaks from $Bi_2Se_3$ (130, 173 cm$^{-1}$) (Figure 1c). The whole film grows in a nearly quintuple layer-by-layer (QL-by-QL) fashion.[14,20,21] In fact, a cross-sectional transmission electron microscopy (TEM) reveals the feature of layered film structure (Figure 1d). In the initial stage of growth, the first quintuple layer (QL) is preferentially nucleated on a graphene surface,[19] and the second QL starts to nucleate before completion of the first QL. To investigate the film growth in detail, particularly, the interface with underlying graphene, we measured X-ray intensity at an anti-Bragg point of $Bi_2Se_3$ during the growth under ultrahigh vacuum condition (Figure 2a), where the waves diffracted from the two adjacent QLs are out-of-phase and thus cancel each other.[22] One quickly notes that the intensity oscillation with a double-QL period indicates a QL-by-QL growth mode. To quantify the data, we have simulated the intensity oscillation based on the atomic model with gradually increasing layer occupancies. We notice here that a few top-layers of the film grow with nearly constant roughness in the initial stage (1–1.5 nm; Figure 2b). Therefore, our film can be qualitatively considered as a good layered material.

In Figure 2c, we observe that the three Bragg peaks (003), (006), and (009) become sharper with increasing thickness and the peak from MLG becomes slightly modified by the interference effect with the film. Note also that the diffraction profiles from the films of 4 QLs and 6 QLs are well described by simulations (dotted curves; see Supporting Information). From the best fits, we find that the interfacial layers between graphene and the first Se are separated by 3.14 Å using MLG as a phase reference. This is a reasonable value compared with the mean value (2.97 Å) of the van der Waals gaps of graphene (3.35 Å) and Se–Se (2.58 Å). From the (006) peak positions, the lattice parameter appears to approach the bulk value even at the first QL and becomes almost completely relaxed in the thicker films (Figure 2d). The in-plane reflections (110) of Bi$_2$Se$_3$ film, MLG, SiC substrate in Figure 2e exhibit relatively well-aligned crystalline orientations.[18,20] Inset in Figure 2e shows the most possible arrangement of the first atomic layer of Se on the graphene lattice considering the lattice match between two different layers. It seems quite surprising that the overlayer lattice is lined up so sharply even in the negligible coupling with the underlying graphene substrates.[23] We, therefore, understand that the film grows in the unique form of QL structure even from the early stage of growth on graphene without being deformed by interfacial strain.

We now discuss our magnetotransport measurements from a mesoscopic device with a Bi$_2$Se$_3$ film (t ~ 400 nm in thickness) grown on a single-layer graphene (SLG/SiO$_2$) as seen in Figure 1.[24] The temperature dependence of resistance shows a weakly metallic behavior, although there is a slight upturn with decreasing temperature below ~20 K (lower inset in Figure 3a).[25,26] Magnetoresistance and Hall resistance ($R_{xx}$, $R_{xy}$) exhibit apparently a nonlinear dependence in the low field region of $B < 3$ T, while weak quantum oscillations are mixed together with linear resistance in high field region for $B > 6.5$ T. In the high field limit, the total carrier density is determined as $n = B/eR_{xy}t$ ~ $1.1 \times 10^{19}$ cm$^{-3}$, where bulk electrons are the most dominant carriers.[25] The nonlinear field dependence of $R_{xy}$ at low magnetic fields, however, suggests the presence of additional conducting channels with higher mobilities $\mu_i$.[27–30] In order to figure out contributions from all possible electrical channels such as bulk, surface, and graphene, we have fitted both resistances ($R_{xx}$, $R_{xy}$) together with the same set of parameters ($n_i$, $\mu_i$; see Supporting Information).

Our fit-curve in Figure 3b shows that the addition of hole channel significantly improves the fitting of multiple carriers. We first evaluate the mobilities of three carriers as 0.61 m$^2$/Vs for the holes, and 0.22, 0.04 m$^2$/Vs for two electrons, from our fit ($h+e_\text{b}+e_\text{im}$). We attribute the hole carriers of high mobility to graphene, and two electrons to $Bi_2Se_3$. The dominant electrons of mobility 0.22 m$^2$/Vs belong to the bulk conduction band, while the lowest mobility electrons to the impurity bands originating from intrinsic defects as reported earlier.[29,30] By adding one more electron channel near the TI–graphene interface ($h+e_\text{b}+e_\text{im}+e_\text{s}$; blue curve), the fitting has been apparently improved, but no further with additional surface channel. We conclude one hole and three electron channels are the optimum number of channels for our TI sample grown epitaxially on graphene substrate. The bulk conduction appears dominant by over an order of magnitude.

We also present the temperature dependence of carrier density $n_i$, mobility $\mu_i$, and conductance $G_i$ of each channel (Figure 3c–e). We notice an order of magnitude smaller carrier density for surface ($n_\text{s} \sim 1\times10^{13}$ cm$^{-2}$) and graphene channels ($n_\text{gr} \sim 8\times10^{12}$ cm$^{-2}$) compared to that of bulk ($n_\text{b} \sim 3.7\times10^{14}$ cm$^{-2}$). The mobility of the surface channel is found to vary sensitively with temperature unlike the other channels, decreasing from 0.5 m$^2$/Vs. In the fitting with four channels, the channel parameters of the surface state reach the optimum values only below 40 K. Considering the very large error bars at higher temperatures for an optimum fit above 40 K, we took the root-mean-square value of the errors as an indication of the interparameter correlation to be minimized together with the overall residual in the fitting procedure. We ascribe this decrease of the mobility to the scatterings of the surface electrons with interfacial phonons in the underlying graphene and interface oxide in addition to the bulk acoustic phonon. However, more detailed investigation is required for clearer understanding.[28,31] The relatively weak but unique surface conduction has been observed consistently by scrutinizing SdH oscillations.

We present the SdH oscillations $\Delta R_{xy}$ for different temperatures in Figure 4a. Although similar oscillating features are also observed from $\Delta R_{xx}$, we focus here on $\Delta R_{xy}$ because of its relatively higher sensitivity to quantum oscillations (see Supporting Information).[29] As expected for a multiple carrier system, we find more than two characteristic SdH frequencies, *i.e.,* two prominent frequencies $F_{SdH}$ = 75 and 156 T, and one very weak frequency $F_{SdH}$ = 425 T in our fast Fourier transformation (FFT) shown in Figure 4b. In another way, the second derivative of $R_{xy}$, -d$^2 R_{xy}$/d$B^2$, provides rather enhanced oscillations showing the same features as those in FFT of $\Delta R_{xy}$ calculated after subtraction of the positive magnetoresistance background. We thus confirm no artifacts associated with the background subtraction.

To determine the origin of these oscillations, we examine only the SdH frequencies corresponding to the three channels, except for the impurity bands, having high mobilities $\mu \sim 1/B_c > 0.15$ m$^2$/Vs obtained above. The hole carriers associated with Dirac-type carriers of graphene may be observed at the corresponding frequency $F_{SdH}$ = $(h/4e)n_{gr} \sim 83$ T.[24] For the bulk conduction band with its Fermi surface size $k_F \sim 0.067$ Å$^{-1}$, together with its carrier density $n_{3D} = n_b/t \sim 1\times 10^{19}$ cm$^{-3}$, one expects an oscillation at $F_{SdH} = (h/4\pi e) k_F^2 \sim 147$ T. These values observed are indeed in good agreement with previous ARPES data reported for Bi$_2$Se$_3$ with high carrier concentration.[25] The observed oscillation at $F_{SdH}$ = 75 T originates from the Dirac Fermi surface of the graphene, while the dominant oscillation of $F_{SdH}$ = 156 T from bulk band electrons. Therefore, we understand that the majority carriers causing the SdH oscillations are the bulk carriers with cyclotron mass $m_c = 0.125 m_e$ (see Supporting Information).

Now the calculated SdH curve with electrons from bulk and holes from graphene fits well the oscillation of $T$ = 15 K (Figure 4a) but deviates from that of $T$ = 2 K. In fact, we notice the presence of an additional frequency associated with the surface state by some distortions of the SdH oscillation of 2 K (marked by arrows). For the fit of SdH curves, we used the phase shifts of γ = 0.2, 0.7, 0.7 for bulk, graphene, surface, respectively, which deviate overall by ~0.2 from the expected Berry phase (γ).[9,13,29] Considering the phase difference (1/4) of $\Delta R_{xx}$ and $\Delta R_{xy}$, an SdH oscillation is described by $\Delta R_{xy} \sim A(T) D(\tau_q) \cos[2\pi(F_{SdH}/B+1/2+\gamma-1/4)]$. The oscillation amplitude is given by $A(T) \sim (2\pi^2 k_B T/\hbar\omega_c) / \sinh(2\pi^2 k_B T/\hbar\omega_c)$, where $\omega_c = eB/m_c$. The Dingle term is $D(\tau_q) \sim \exp(-\pi/\omega_c\tau_q)$, where $\tau_q$ is a quantum scattering time. However, an exact value of the Berry phase is not determined due to our data obtained in a limited range of magnetic fields as well as an overall phase deviation in $\Delta R_{xy}$. The surface channel causing the unique behaviour of conductance shown in Figure 3 may well be attributed to another frequency $F_{SdH}$ = 425 T observed in FFT. Considering the nondegeneracy of a topological metallic state, we estimate $F_{SdH} = (h/e)n_s \sim 415$ T for the surface channel. As stated above, the fastest one of $F_{SdH}$ = 425 T manifests the existence of another Fermi surface associated with the surface electrons of the TI.[26] It is also interesting to note that this fast oscillation disappears at higher temperature above ~10 K.[13] We can also estimate the $m_c$ of the surface electrons from the temperature dependence of FFT amplitudes at $F_{SdH}$ = 425 T. The FFT amplitudes should have the same temperature dependence as $A(T)$ which decreases gradually with increasing temperature, since $A(T)$ is nearly constant within the range of experimental magnetic fields. We thus obtained $m_c \sim 0.2 m_e$ for the surface state, which produces the $\tau_q = (m_c/e)\mu \sim 5\times10^{-13}$ s an order of magnitude larger than that of bulk carriers. Here, we exclude other possibility of two-dimensional electron gas (2DEG) formed by surface band bending,[25,32–34] which may be seen around $F_{SdH} \sim 210$ T representing smaller Fermi surface with $n_s$, since no such a peak for 2DEG appears in our FFT. We have thus reproduced almost the same features of overall oscillations by adding three oscillatory functions with different frequencies. Therefore, our data for the quantum oscillations in $Bi_2Se_3$ film unambiguously reveal the presence of the topological surface state in the buried interface.

**CONCLUSION**

We have grown the epitaxial $Bi_2Se_3$ film on graphene and investigated the interfacial structure in the atomic level by an *in situ* X-ray diffraction together with its electrical conductance by measuring the SdH oscillations. We find an epitaxial growth of the film in a QL-by-QL mode preserving the surface state nearly intact with no interfacial strain effect. Out of the four possible conductance channels we identified from the transport measurements, the contribution to the total conduction from surface electrons of the TI's surface states manifests itself in both mobility and conduction by showing a distict temperature dependence. We also notice the presence of a distinct but quite weak surface signal in FFT with a SdH frequency much higher than the other two dominant frequencies from bulk and graphene. This is the signal from the surface states of TI persistent in the graphene-TI hybrid system, distinguished from other channels in conductivity and also the SdH oscillation of the weak surface signal is quite different from that of graphene or of the bulk. We thus provide a pragmatic possibility to utilize surface current of TI materials, weak as it is, in the novel graphene-supported topological insulator devices with enhanced performance by forming high-quality interfaces.[16]

**EXPERIMENTAL SECTION**

***In situ* X-ray diffraction**     We have prepared different types of graphene samples:[17] (i) CVD graphene on SiO$_2$ and (ii) multilayer graphene (MLG) and single-layer graphene (SLG) grown on 6H–SiC(0001) surface. The CVD graphene grown on a copper foil has been transferred to the SiO$_2$ substrate after an etching process. MLG was grown in an ultrahigh vacuum (UHV) chamber following the growth recipe. The samples were mounted in an *in situ* X-ray scattering chamber working under UHV, $< 1\times10^{-9}$ Torr. Bi$_2$Se$_3$ pieces wrapped with a tungsten wire were placed separately ~10 cm apart from the sample substrate. The Bi$_2$Se$_3$ was thermally evaporated by heating the tungsten wire resistively. Film growth has been performed typically at an effective growth rate ($\leq 0.05$ QL/min) with the substrate maintained near 300 °C. Measurements were performed at the 5D BL of Pohang Light Source for *in situ*, and at the 5A BL for *ex situ*.

**Electrical transport**     Single-layer graphene (SLG), characterized by Raman measurements (532 nm), has been placed by mechanical exfoliation onto 300-nm thick SiO$_2$/Si using adhesive tape.[24] A Hall-bar shaped device was fabricated in the standard electron beam lithography followed by the lift-off process. Electrical contacts, Ti/Au (5/15 nm), were made on SLG/SiO$_2$ before Bi$_2$Se$_3$ deposition. The device sample was put in a quartz tube together with Bi$_2$Se$_3$ crystal pieces placed ~10 cm apart. The pressure of the tube was maintained below $10^{-5}$ Torr during the growth by pumping through the sample side of the tube. Bi$_2$Se$_3$ was thermally evaporated above 450 °C in a furnace and it was deposited for 3 days on the SLG/SiO$_2$ device sample heated up to 300 °C. The film thickness was measured by using AFM and optical profiler (Wyko NT1100). Four-probe measurements have been performed by standard low-frequency lock-in techniques with a current bias of $I = 1 \sim 5$ μA using a physical property measurement system (Quantum Design, PPMS). Magnetotransport data were measured under a magnetic field up to 14 T applied along the direction normal to the sample surface.

**Supporting Information.** Atomic structure of $Bi_2Se_3$ on graphene and device structure; fitting of diffraction intensity profiles; X-ray reflectivity; AFM images for $Bi_2Se_3$ film on SiC, film on CVD graphene/$SiO_2$, film on microscale exfoliated graphene/$SiO_2$; conductance of each channel as a function of $B$; $T$-, $B$-dependence of SdH oscillations; comparison of $\Delta R_{xx}$ ($\Delta R_{xy}$) SdH oscillations; analysis of the magnetotransport for multiple carriers.

**Acknowledgment**

We thank H. W. Lee, M. Koshino, and P. Kim for helpful suggestions on improvements.

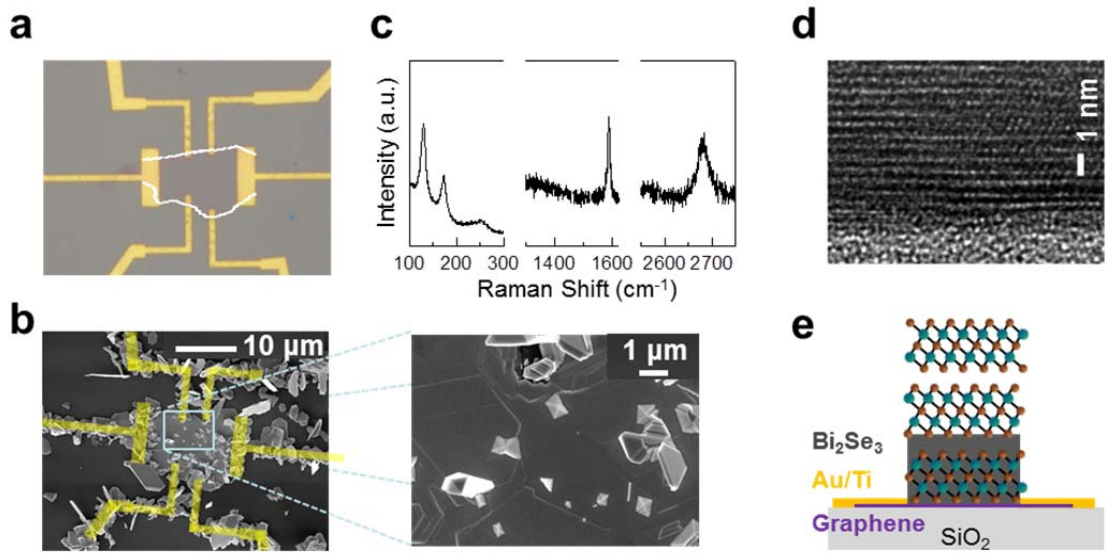

**Figure 1.** $Bi_2Se_3$ film growth on graphene. (a) Optical image of a graphene device with 20-nm-thick electrical leads (Au/Ti) on $SiO_2$ substrate. (b) SEM image taken after $Bi_2Se_3$ film growth (400 nm) on the graphene device. A magnified image shows the surface morphology. (c) Raman spectra showing the characteristic peaks of $Bi_2Se_3$ and SLG. (d) A cross-sectional TEM image of the device sample showing a layered film stacked on graphene/$SiO_2$. (e) A schematic of the device structure.

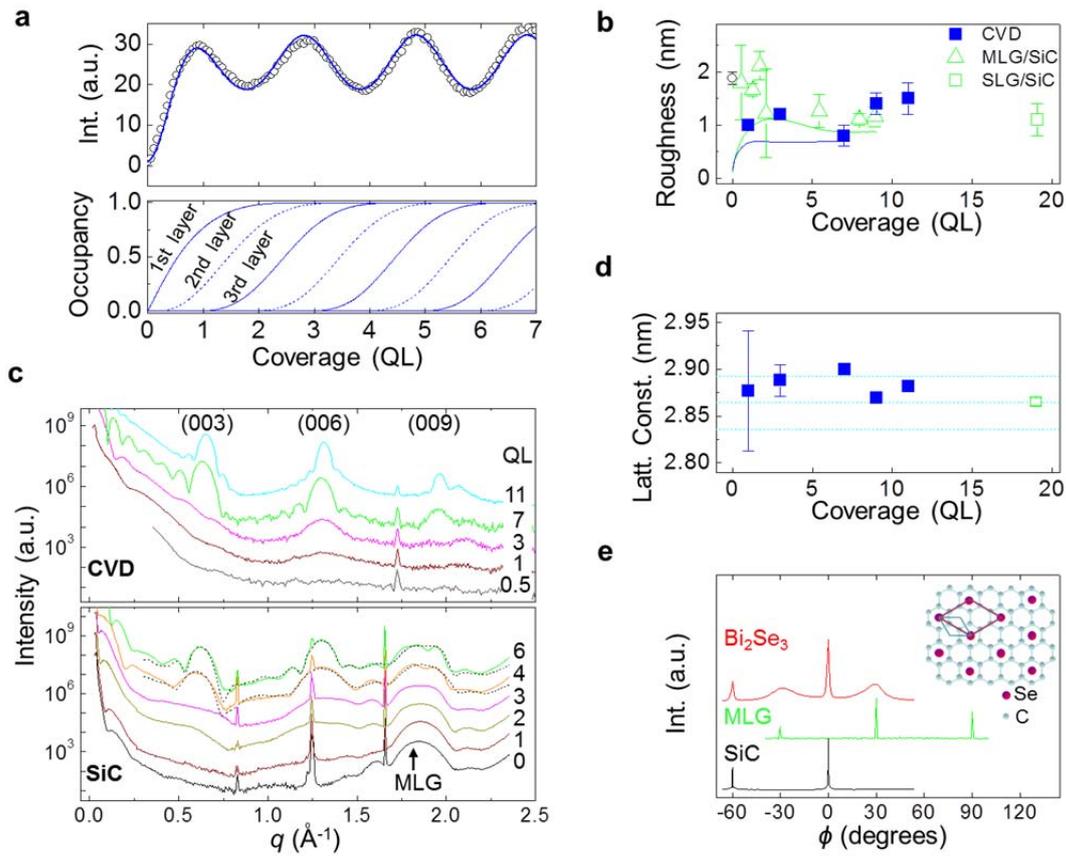

**Figure 2.** *In situ* X-ray diffraction. (a) Change of intensity oscillation observed at an anti-Bragg point (0 0 3/2) of the film during the growth of the TI film on CVD graphene, showing a QL-by-QL growth mode. The solid curves are simulations based on the atomic model considering the varying layer occupancy shown in the lower panel. (b) Roughness obtained from reflectivity (blue square for the film on CVD, green triangle for the film on MLG/SiC, green square for the film on SLG/SiC, and black circle for MLG/SiC alone). Solid curves (blue for the film on CVD, green for the film on MLG/SiC) represent the lower bounds of roughness estimated with the layer occupancies obtained from the intensity oscillation observed at the anti-Bragg point. (c) Diffraction profiles with increasing film thickness as a function of perpendicular momentum transfer $q$. Progressive change of the Bragg peaks (003), (006), and (009) of the $Bi_2Se_3$ film grown on CVD graphene (upper). For the growth on graphene/SiC, the Bragg reflection of MLG and its interference with the film are observed (lower). The films of 4 QLs and 6 QLs are well described by the simulations based on the kinematic calculation (dotted curves). (d) Lattice parameter obtained from the (006) Bragg peak where the bulk value is indicated by cyan dotted lines with its upper and lower bounds within one percent offset. (e) In-plane reflections (110) of the $Bi_2Se_3$ film, MLG, and SiC substrate showing relatively well-aligned crystalline orientations. The first atomic layer of Se (purple circles) is arranged on the graphene lattice.

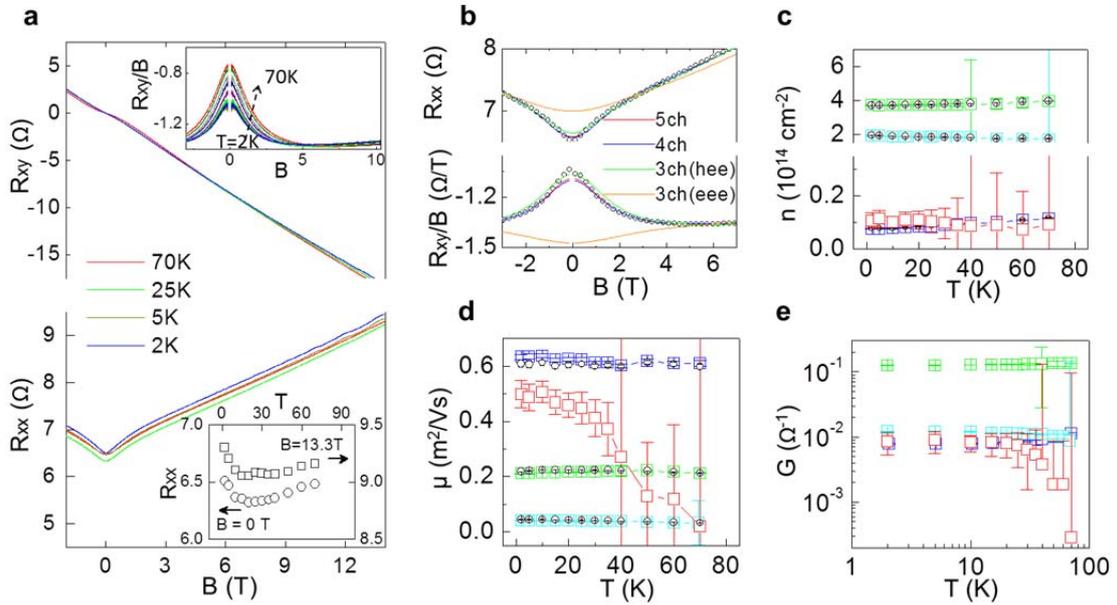

**Figure 3.** Magnetotransport measurements. (a) Changes of magnetoresistance ($R_{xx}$) and Hall resistance ($R_{xy}$) with increasing magnetic field $B$ at several different temperatures ranging from 2 to 70 K. Upper inset shows two-dimensional Hall coefficient ($R_{xy}/B$) as a function of $B$, while lower inset does the temperature dependence of $R_{xx}$ for $B = 0$ and 13.3 T. (b) Fit curves of $R_{xx}$ and $R_{xy}$ with varying number of possible conductance channels, where the best fit is found with four different channels (blue curve). Data of $R_{xx}$, $R_{xy}/B$ at $T = 2$ K (black open circles). (c),(d) Temperature dependence of each carrier density $n_i$, and mobility $\mu_i$. Each $n_i$ and $\mu_i$ were obtained from the best fit with the four different channels (open squares: blue for graphene, red for surface channel, cyan for impurity bands, and green for bulk conduction band). Small open circles (black) denote the fitting values done with three channels excluding surface channels. (e) Temperature dependence of each conductance.

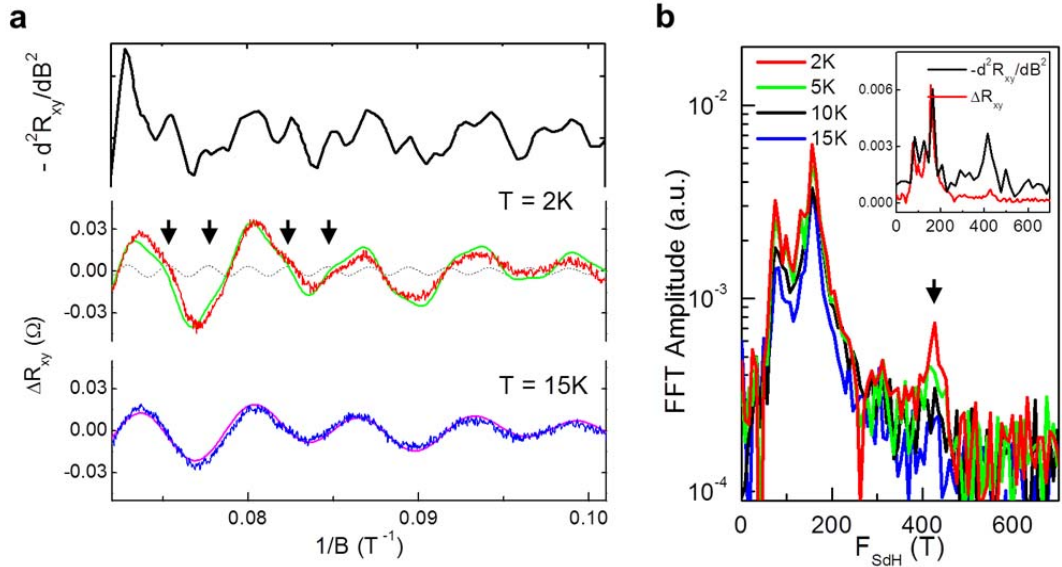

**Figure 4.** Shubnikov-de Haas (SdH) oscillations. (a) SdH oscillations ($\Delta R_{xy}$) of 2 K (red) and 15 K (blue). Simulations for the experimental SdH oscillations of 2 K with three frequencies (green curve) and of 15 K with two channels (pink). Arrows indicate distortions away from a smooth oscillating curve caused by the surface channel of frequency $F_{SdH} \sim 425$ T (dotted curve). The derivative of $R_{xy}$, $-d^2R_{xy}/dB^2$, exhibits rather enhanced oscillations. (b) Fast Fourier transformation (FFT) of $\Delta R_{xy}$ data at four different temperatures showing three peaks of different origin. Two prominent frequencies $F_{SdH} \sim 75, 156$ T originate from the hole carriers of graphene and the bulk band electrons, respectively. For clarity of the weak peak $F_{SdH} \sim 425$ T (down arrow) originating from the TI's surface channel, our FFT of the derivative of $R_{xy}$, $-d^2R_{xy}/dB^2$, reproduces the same peaks (inset).

**Table of Contents**

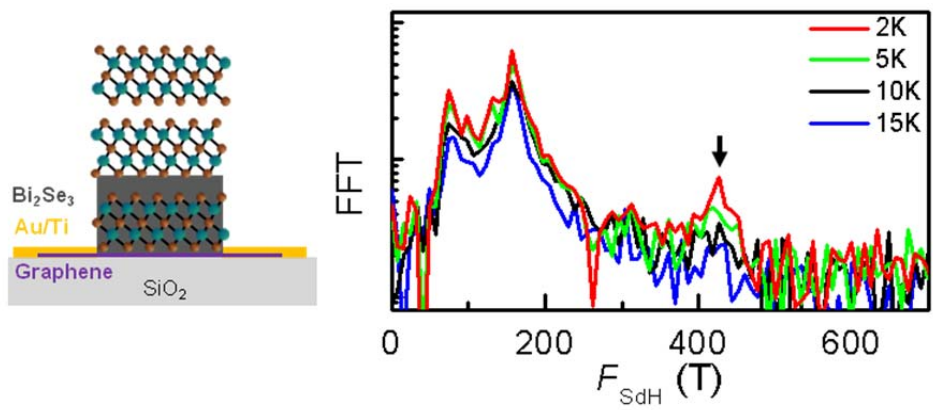